\documentclass[8.5pt,twoside,twocolumn]{article}
\usepackage[super,sort&compress,comma]{natbib}
\usepackage[version=3]{mhchem}
\usepackage{times}
\usepackage{sectsty}
\usepackage{balance}

\usepackage{graphicx} %eps figures can be used instead
\usepackage{lastpage}
\usepackage[format=plain,justification=raggedright,singlelinecheck=false,font=small,labelfont=bf,labelsep=space]{caption}
\usepackage{fancyhdr}
\newcommand{\matrixel}[3]{%
    \langle #1 | #2 | #3 \rangle}
\newcommand{\epratio}[1]{%
    $m_p/m_e\:$}
\newcommand{\dpol}[1]{%
    \alpha_{vJM}(\omega)}
\newcommand{\dpolAv}[1]{%
    \overline {\alpha_{vJ}(\omega)}}
\newcommand{\dpolO}[1]{%
    \alpha_{vJM}(0)}
\newcommand{\dpole}[1]{%
    \alpha^e_{vJM}(\omega)}
\newcommand{\dpoleAv}[1]{%
    \overline{\alpha^e_{vJ}(\omega)}}
\newcommand{\dpoleO}[1]{%
    \alpha^e_{vJM}(0)}
\newcommand{\dpolrv}[1]{%
    \alpha^{rv}_{vJM}(\omega)}
\newcommand{\dpolrvAv}[1]{%
    \overline{\alpha^{rv}_{vJ}(\omega)}}
\newcommand{\dpolrvO}[1]{%
    \alpha^{rv}_{vJM}(0)}

\begin{document}

%\thispagestyle{plain}
%\fancypagestyle{plain}{
%\fancyhead[L]{\includegraphics[height=8pt]{headers/LH}}
%\fancyhead[C]{\hspace{-1cm}\includegraphics[height=20pt]{headers/CH}}
%\fancyhead[R]{\includegraphics[height=10pt]{headers/RH}\vspace{-0.2cm}}
\fancypagestyle{plain}{
\renewcommand{\headrulewidth}{1pt}}
\renewcommand{\thefootnote}{\fnsymbol{footnote}}
\renewcommand\footnoterule{\vspace*{1pt}%
\hrule width 3.4in height 0.4pt \vspace*{5pt}}
\setcounter{secnumdepth}{5}

\makeatletter
\def\subsubsection{\@startsection{subsubsection}{3}{10pt}{-1.25ex plus -1ex minus -.1ex}{0ex plus 0ex}{\normalsize\bf}}
\def\paragraph{\@startsection{paragraph}{4}{10pt}{-1.25ex plus -1ex minus -.1ex}{0ex plus 0ex}{\normalsize\textit}}
\renewcommand\@biblabel[1]{#1}
\renewcommand\@makefntext[1]%
{\noindent\makebox[0pt][r]{\@thefnmark\,}#1}
\makeatother
\renewcommand{\figurename}{\small{Fig.}~}
\sectionfont{\large}
\subsectionfont{\normalsize}

\fancyfoot{}
%\fancyfoot[LO,RE]{\vspace{-7pt}\includegraphics[height=9pt]{headers/LF}}
%\fancyfoot[CO]{\vspace{-7.2pt}\hspace{12.2cm}\includegraphics{headers/RF}}
%\fancyfoot[CE]{\vspace{-7.5pt}\hspace{-13.5cm}\includegraphics{headers/RF}}
\fancyfoot[RO]{\footnotesize{\sffamily{1--\pageref{LastPage} ~\textbar  \hspace{2pt}\thepage}}}
\fancyfoot[LE]{\footnotesize{\sffamily{\thepage~\textbar\hspace{3.45cm} 1--\pageref{LastPage}}}}
\fancyhead{}
\renewcommand{\headrulewidth}{1pt}
\renewcommand{\footrulewidth}{1pt}
\setlength{\arrayrulewidth}{1pt}
\setlength{\columnsep}{6.5mm}
\setlength\bibsep{1pt}

\twocolumn[
  \begin{@twocolumnfalse}
\noindent\LARGE{\textbf{Infrared dynamic polarizability of HD$^+$ rovibrational states}}%$^\dag$}}
\vspace{0.6cm}

\noindent\large{\textbf{J.C.J. Koelemeij$^{\ast, \dag}$}}\vspace{0.5cm}
%Please note that \ast indicates the corresponding author(s) but no footnote text is required.

%\noindent\textit{\small{\textbf{Received Xth XXXXXXXXXX 20XX, Accepted Xth XXXXXXXXX 20XX\newline
%First published on the web Xth XXXXXXXXXX 200X}}}
%
%\noindent \textbf{\small{DOI: 10.1039/b000000x}}
\vspace{0.6cm}
%Please do not change this text.

\noindent \normalsize{A calculation of dynamic polarizabilities of rovibrational states with vibrational quantum number $v=0-7$ and rotational quantum number $J=0,1$ in the 1s$\sigma_g$ ground-state potential of HD$^+$ is presented. Polarizability contributions by transitions involving other 1s$\sigma_g$ rovibrational states are explicitly calculated, whereas contributions by electronic transitions are treated quasi-statically and partially derived from existing data [R.E. Moss and L. Valenzano, \textit{Molec. Phys.}, 2002, \textbf{100}, 1527]. Our model is valid for wavelengths $>4~\mu$m and is used to to assess level shifts due to the blackbody radiation (BBR) electric field encountered in experimental high-resolution laser spectroscopy of trapped HD$^+$ ions.   Polarizabilities of 1s$\sigma_g$ rovibrational states obtained here agree with available existing accurate \textit{ab initio} results. It is shown that the Stark effect due to BBR is dynamic and cannot be treated quasi-statically, as is often done in the case of atomic ions. Furthermore it is pointed out that the dynamic Stark shifts have tensorial character and depend strongly on the polarization state of the electric field. Numerical results of BBR-induced Stark shifts are presented, showing that Lamb-Dicke spectroscopy of narrow vibrational optical lines ($\sim 10$ Hz natural linewidth) in HD$^+$ will become affected by BBR shifts only at the $10^{-16}$ level. }
\vspace{0.5cm}
 \end{@twocolumnfalse}
  ]

\section{Introduction}\label{intro}
%Footnotes
%\footnotetext{\dag~Electronic Supplementary Information (ESI) available: [details of any supplementary information available should be included here]. See DOI: 10.1039/b000000x/}

%Please use \dag to cite the ESI in the main text of the article.
%If you article does not have ESI please remove the the \dag symbol from the title and the above footnotetext.

\footnotetext{\textit{$^{\ast}$~LaserLaB, Vrije Universiteit, De Boelelaan 1081, 1081 HV Amsterdam, Netherlands. Fax: +31 (0)20 598 7992; Tel: +31 (0)20 589 7903; E-mail: koel@few.vu.nl}}
%$\footnotetext{\textit{$^{b}$~Address, Address, Town, Country. }}

%additional addresses can be cited as above using the lower-case letters, c, d, e... If all authors are from the same address, no letter is required

\footnotetext{$^{\dag}$~Acknowledges the Netherlands Organisation for Scientific Research for support.}
The molecular hydrogen ion (H$_2^+$) and its isotopomers (HD$^+$, D$_2^+$, HT$^+$, \textit{etc.}) are the simplest naturally occurring molecules. As such they are amenable to high-accuracy \textit{ab initio} level structure calculations, which are currently approaching 0.1~ppb for rovibrational levels in the electronic ground potential\cite{Korobov2008}. The inclusion of high-order QED terms in these calculations makes molecular hydrogen ions an attractive subject for experiments aimed at comparison with theory and tests of QED. With rovibrational states having lifetimes exceeding 10~ms it has long been recognized that optical (infrared) spectroscopy could provide accurate experimental input, and several experimental studies were undertaken \cite{Wing1976long,Koelemeij2007a} or are currently in progress \cite{Karr2011}. The highest accuracy that has hitherto been achieved is 2~ppb for a Doppler-broadened vibrational overtone transition at $1.4~\mu$m in trapped HD$^+$ molecular ions, sympathetically cooled to 50~mK \cite{Koelemeij2007a}. By comparison, the highest accuracy achieved in laser spectroscopy of laser-cooled atomic ions, tightly confined in the optical Lamb-Dicke regime, is $\sim 1 \times 10^{-17}$  in the case of the Al$^+$ optical clock at NIST Boulder, USA.\cite{Chou2010a}. The Al$^+$ optical clock employs quantum-logic spectroscopy (QLS) which utilizes entangled quantum states of two trapped ions, one of which is used for (ground-state) laser cooling and efficient state detection, whereas the other ion contains the transitions of spectroscopic interest\cite{Schmidt2005}. It has been pointed that Doppler-free spectroscopy may be performed on HD$^+$ as well\cite{Koelemeij2007a}, and also that QLS may be used for spectroscopy of molecular ions\cite{Schmidt2006long}.
\begin{figure}[!ht]
  \centering
  \includegraphics[height=7.5cm]{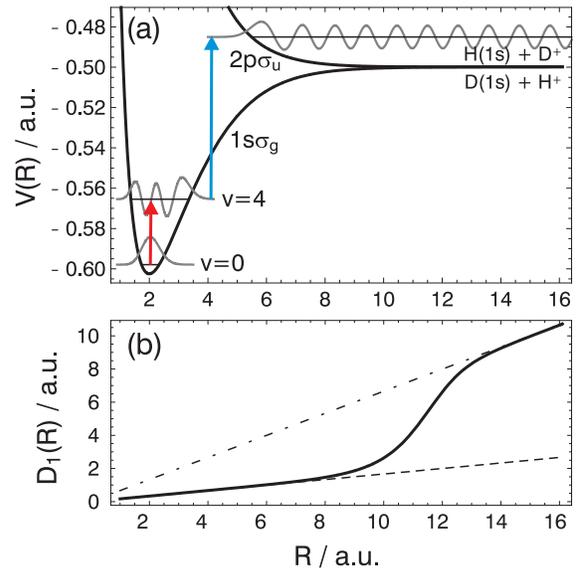}
  \caption{(Color online)(a) Potential energy curves of the 1s$\sigma_g$  and 2p$\sigma_u$ electronic sates. Indicated energy values are binding energies of the molecule. Shown also are radial nuclear (vibrational) wavefunctions, $\chi_{v}(R)$, for $v=0,4$ as well as one dissociating nuclear wavefunction in the 2p$\sigma_u$ state. The red arrow represents a purely rovibrational transition within 1s$\sigma_g$; the blue arrow exemplifies a transition between different electronic states. (b) Dipole moment function $D_1(R)$ used in the calculation of radial dipole matrix elements (solid curve), shown together with the approximate function used by Colbourn and Bunker\cite{Colbourn1976} (dashed line) and the fully \textit{g/u} symmetry-broken dipole moment function, valid at long internuclear range (dot-dashed line).}
  \label{fig:pots}
\end{figure}

Accurate results of laser spectroscopy of HD$^+$ are of interest for the determination of the value of the proton-electron mass ratio, \epratio\ \cite{Wing1976long}, and for the search for a variation of \epratio\ with time\cite{Schiller2005}. The former may be achieved by combining \textit {ab initio} theoretical results with results from spectroscopy at an accuracy level of $\sim 10^{-10}$; for the latter spectroscopic results with an accuracy of $\sim 10^{-15}$ are required to improve on the current most stringent bounds\cite{Klein2007,Blatt2008}. In both cases spectroscopy of optical transitions is faced with level shifts due to magnetic and electric fields and, to a lesser extent, shifts due to collisions and relativistic effects. The Zeeman effect of HD$^+$ was recently considered by Bakalov \textit{et al.}, and level shifts to second order in the magnetic field were given for a large set of rovibrational states\cite{Bakalov2010, Bakalov2011}. Static polarizabilities of vibrational states with rotational quantum number $J=0,1$ were calculated and reported by several authors\cite{Moss2002,Karr2005}, while dynamic polarizabilities of HD$^+$ vibrational states with $J=0$ were evaluated for a discrete set of two-photon transition wavelengths in the 1--18~$\mu$m wavelength range\cite{Karr2005}. However, to our knowledge, no results on dynamic polarizabilities of HD$^+$ for vibrational states with $J>0$ are available in literature. Polarizabilities of such states for a wide range of infrared wavelengths are required for the calculation of differential Stark shifts due to blackbody radiation (BBR). Moreover, since the BBR spectrum encompasses several rovibrational transitions of the HD$^+$ ion, it is expected that the quasi-static treatment of BBR-induced Stark shifts as often done in the case of atomic ion species is not valid for HD$^+$. Rather, the case of HD$^+$ will be analogous to that of neutral polar molecules, for which BBR-induced Stark shifts were evaluated using dynamic polarizabilities\cite{Vanhaecke2007}.

This Article addresses the (BBR-induced) dynamic Stark effect of HD$^+$ and is organized as follows. In Sec.~\ref{theo} we present our model to calculate dynamic polarizabilities and BBR-induced Stark shifts, followed by a discussion of the results for several rovibrational states in the 1s$\sigma_g$ ground-state potential of HD$^+$ in Sec.~\ref{res}. Conclusions are presented in Sec.~\ref{conc}. Throughout this Article, the terms 'Stark effect' and 'polarizability' will be used interchangeably, and SI units will be used.
\section{Theory}\label{theo}
Figure~\ref{fig:pots}(a) shows a partial energy level diagram of the HD$^+$ molecular ion including the electronic ground-state potential, 1s$\sigma_g$, and the first electronically excited potential, 2p$\sigma_u$. Note that in HD$^+$ the \textit{g/u} symmetry quantum labels are only approximately good quantum labels as the nonidentical nuclei introduce \textit{g/u} symmetry breaking at large internuclear range. The potential energy curves shown are interpolations of data published by Esry and Sadeghpour\cite{Esry1999}, who present the potential energy as the sum of a nonrelativistic, fully adiabatic curve, and a diagonal nonadiabatic correction. We use these curves to obtain (real-valued) radial wavefunctions of nuclear motion, $\chi_{vJ}(R)$, by numerical solution of the radial Schr\"odinger equation including the centrifugal term due to the molecular rotation:
\begin{equation}\label{eq:RadSchrod}
 - \frac{\hbar^2}{2\mu} \frac{\mathrm{d}^2}{\mathrm{d} R^2}\chi_{vJ}(R) + \left[ V_i(R) +\frac{\hbar^2 J(J+1)}{2 \mu R^2} \right] \chi_{vJ}(R) = E_{vJ}\chi_{vJ}(R),
\end{equation}
where $R$ denotes the internuclear separation, $\mu$ stands for the nuclear reduced mass of the molecule, $v$ labels the vibrational state, $J$ is the rotational angular momentum of the molecule, and $E_{v,J}$ is the rovibrational energy. $V_i(R)$ are the potential energy curves for the $i=$ 1s$\sigma_g$, 2p$\sigma_u$ states taken from Esry and Sadeghpour\cite{Esry1999}, who also provide dipole moment functions $D_1(R)$ and $D_{12}(R)$. These correspond to the dipole moment of the 1s$\sigma_g$ state and the dipole moment of electronic transitions between 1s$\sigma_g$ and 2p$\sigma_u$, respectively.

The dynamic polarizability corresponds to the ability of the HD$^+$ molecule to deform under the influence of an oscillating electric field, and depends on the strengths and frequencies of many electric dipole transitions in both the nuclear and the electronic degrees of freedom. Laser spectroscopy on HD$^+$ is typically performed on transitions between low-lying rovibrational levels  in the 1s$\sigma_g$ state, and it is the dynamic polarizability of these levels that we will focus on here. The dynamic polarizability, $\alpha(\omega)$, is defined as follows. A quantum state with quantum numbers $(v,J,M)$ (with $M$ corresponding to the projection of $\mathbf{J}$ on the space-fixed $z$-axis) and energy $E_{vJM}$ will undergo an energy shift $\Delta E$ due to the interaction with a monochromatic electric field with amplitude $\mathcal{E}$, polarization state $q$, and angular frequency $\omega$ equal to
\begin{equation}\label{eq:dynpol}
\Delta E = - \frac{1}{4} \alpha_{vJM}^q(\omega) \mathrm{\mathcal{E}} ^2
\end{equation}
In the remainder of this Article, the polarization state $q\in (-1,0,1)$ will be taken $q=0$ (\textit{i.e.} linear polarization parallel to the space-fixed $z$-axis) and we will omit the label $q$ altogether; see Sec.~\ref{theo:rovib}. The dynamic polarizability can be written as $\dpol\ = \dpolrv\ + \dpole\ $, where $\dpolrv\ $ stands for the contribution by transitions coupling to other 1s$\sigma_g$ rovibrational states, and $\dpole\ $ accounts for the contributions by all transitions connecting to electronically excited states. To simplify the calculation, we will restrict ourselves to the two strongest sets of transitions from the rovibrational states of interest. These are (1) purely rovibrational transitions within the electronic ground state 1s$\sigma_g$, and (2) electronic dipole transitions to dissociating states in 2p$\sigma_u$.
\begin{table}[!ht]
\small
  \caption{\ Static polarizabilities (in units of $4\pi \epsilon_0 a_0^3$) for vibrational states with $J=0$. Total polarizabilities $\dpolO\ $ were taken from Moss and Valenzano\cite{Moss2002}. Individual rovibrational and electronic contributions $\dpolrvO\ $ and $\dpoleO\ $, respectively, are also specified. Entries in the rightmost column are obtained from those in the other columns as $\dpolO\ - \dpolrvO\ $}
  \label{table:pols}
  \begin{tabular*}{0.5\textwidth}{@{\extracolsep{\fill}}llll}
    \hline
    $v$ & $\dpolO\ $~(Ref.~[14]) & $\dpolrvO\ $ (this work) & $\dpoleO\ $ \\
    \hline
0 & 395.306 & 392.2 & 3.1 \\
1 & 462.65 & 458.9 & 3.8 \\
2 & 540.69 & 536.2 & 4.5 \\
3 & 631.4 & 625.9 & 5.5 \\
4 & 737.3 & 730.7 & 6.6 \\
5 & 861.7 & 853.5 & 8.2 \\
6 & 1008 & 998.6 & 9.4 \\
7 & 1184 & 1171 & 13 \\
    \hline
  \end{tabular*}
\end{table}
\begin{table}[!ht]
\small
  \caption{\ Comparison of purely electronic static polarizabilities $\dpoleO\ $(in units of $4\pi \epsilon_0 a_0^3$) for vibrational states with $J=0$ of HD$^+$ with accurate values for vibrational states with $J=0$ of H$_2^+$ and D$_2^+$, calculated by Hilico \textit{et al.}\cite{Hilico2001} Note that for H$_2^+$ and D$_2^+$ the static polarizability stems from electronic transitions only}
  \label{table:comparison}
  \begin{tabular*}{0.5\textwidth}{@{\extracolsep{\fill}}llll}
    \hline
    $v$ & H$_2^+$\cite{Hilico2001} & HD$^+$ (this work) & D$_2^+$\cite{Hilico2001} \\
    \hline
0 & 3.168\,725\,803 & 3.1 & 3.071\,988\,696 \\
1 & 3.897\,563\,360 & 3.8 & 3.553\,025\,791 \\
2 & 4.821\,500\,365 & 4.5 & 4.119\,581\,678 \\
3 & 6.009\,327\,479 & 5.5 & 4.791\,282\,711 \\
4 & 7.560\,453\,090 & 6.6 & 5.593\,314\,877 \\
5 & 9.621\,773\,445 & 8.2 & 6.558\,318\,701 \\
6 & 12.41\,599\,987 & 9.4 & 7.729\,054\,615 \\
7 & 16.290\,999\,14 & 13  & 9.162\,209\,589 \\
    \hline
  \end{tabular*}
\end{table}
\subsection{Dynamic polarizability due to rovibrational transitions in 1s$\sigma_g$}\label{theo:rovib}
As a starting point we will use the energy shift derived using time-dependent second-order perturbation theory, [Eq.~(7.73)] in the textbook by Sobelman\cite{Sobelman},
\begin{equation}\label{eq:Sobelmandef}
\Delta E_{vJM}(\omega) = \frac{1}{2 \hbar}\mathrm{\mathcal{E}} ^2 \sum_{v',J',M'} \frac{\omega_{vv'JJ'}}{\omega_{vv'JJ'}^2 - \omega^2}|D_{vv'JJ'MM'q}|^2,
\end{equation}
where $v,J,M$ and $v',J',M'$ are the quantum numbers of the initial and final states, respectively. Here we assume that the states $(v,J,M)$ are degenerate in the quantum number $M$. It is furthermore important to note the role of the sign in the definition of $\omega_{vv'JJ'}$:
\begin{equation}\label{eq:omega}
\omega_{vv'JJ'} = ( E_{v'J'}-E_{vJ})/\hbar.
\end{equation}
Hence, $\omega_{vv'JJ'} > 0$ for transitions to more highly-excited states and $\omega_{vv'LL'} < 0$ for transitions to lower states. For purely rovibrational transitions, the squared dipole transition matrix element $|D_{vv'JJ'MM'q}|^2$ reduces to\cite{CarringtonBrown}
%\begin{equation}
%|D_{vv'JJ'MM'}|^2 = |\matrixel{J M}{\cos \theta}{J' M'}|^2 \mu_{vv'JJ'}^2,
%\end{equation}
{\setlength\arraycolsep{2pt}
\begin{eqnarray}\label{eq:robivdipmom}
|D_{vv'JJ'MM'q}|^2
          & = &|\matrixel{J M}{\mathcal{D}^*_{-q0}(\omega_\mathrm{E})}{J' M'}|^2 \mu_{vv'JJ'}^2 \nonumber \\
          & = & (2J+1)(2J'+1)\left( \begin{array}{ccc}
J & 1 & J' \\
0 & 0 & 0
\end{array} \right)^2 \nonumber \\
&&\times
\left( \begin{array}{ccc}
J & 1 & J' \\
-M & -q & M'
\end{array} \right)^2 \mu_{vv'JJ'}^2,
\end{eqnarray}}
with
\begin{equation}\label{eq:radrobivdipmom}
\mu_{vv'JJ'}^2= |\int_{0}^{\infty} \chi_{v'J'}(R)D_1(R) \chi_{vJ}(R)\mathrm{d}R|^2.
\end{equation}
The dipole matrix element $\mu_{vv'JJ'}$ is a vector oriented along the internuclear axis of the HD$^+$ molecule. Therefore, in order to evaluate the matrix elements, $\mu_{vv'JJ'}$ needs to be transformed from the molecule-fixed to the space-fixed frame by rotation about the set of Euler angles $\omega_\mathrm{E}$, which is implemented through the rotation operator $\mathcal{D}^*_{-q0}(\omega_\mathrm{E})$ in the first factor in Eq.~(\ref{eq:robivdipmom}). In arriving at the second line of Eq.~(\ref{eq:robivdipmom}) we use the fact that for states with $\Lambda=0$ (like for 1s$\sigma_g$, while ignoring the spins of the proton, deuteron and electron) the projection of $J$ on the internuclear axis is zero.
As stated in Sec.~\ref{theo}, we will consider the case $q=0$ only.

The squared matrix elements $\mu_{vv'JJ'}^2$ are readily evaluated using the numerical expressions for wavefunctions and dipole moment functions introduced above. The expression for $\dpolrv\ $ is obtained after inserting Eqs.~(\ref{eq:omega}) and (\ref{eq:robivdipmom}) into Eq.~(\ref{eq:Sobelmandef}), followed by equating Eq.~(\ref{eq:Sobelmandef}) to Eq.~(\ref{eq:dynpol}) and solving for $\dpolrv \ $ (momentarily assuming that $\dpole\ = 0$). As we here focus on low-lying vibrational levels and dipole transitions only, we will truncate the summation in Eq.~(\ref{eq:Sobelmandef}) to $v=9$, and also ignore the contribution by purely rovibrational transitions to continuum states above the 1s$\sigma_g$ dissociation limit. This is justified as the line strength of vibrational overtones decreases rapidly with increasing order of the overtone. The summation is furthermore limited to terms obeying the selection rule $J'=J\pm 1$.
\begin{table*}[!ht]
\small
  \caption{\ Static polarizabilities (in units of $4\pi \epsilon_0 a_0^3$) for vibrational states with $J=1$. Total polarizabilities $\dpolO\ $ were taken from Moss and Valenzano\cite{Moss2002}.  Individual rovibrational and electronic contributions $\dpolrvO\ $ and $\dpoleO\ $, respectively, are also specified. For each $M$ value, entries in the rightmost column are obtained from those in the other columns as $\dpolO\ - \dpolrvO\ $}
  \label{table:polsJ1M0}
  \begin{tabular*}{\textwidth}{@{\extracolsep{\fill}}lllllll}
    \hline
     & \multicolumn{3}{c}{$M=0$} & \multicolumn{3}{c}{$|M|=1$}\\
    \cline{2-4} \cline{5-7}
    $v$ & $\dpolO\ $~(Ref.~[14]) & $\dpolrvO\ $ (this work) & $\dpoleO\ $ & $\dpolO\ $~(Ref.~[14]) & $\dpolrvO\ $ (this work) & $\dpoleO\ $ \\
    \hline
0 & -229.986 & -234.1 & 4.2 & 120.979 & 118.4 & 2.6 \\
1 & -268.90 & -274.0 & 5.1 & 141.50 & 138.5 & 3.0 \\
2 & -313.87 & -320.2 & 6.4 & 165.29 & 161.7 & 3.6 \\
3 & -366.00 & -373.9 & 7.9 & 192.95 & 188.8 & 4.2 \\
4 & -426.66 & -436.6 & 9.9 & 225.26 & 220.3 & 4.9 \\
5 & -497.57 & -510.0 & 12 & 263.22 & 257.3 & 5.9 \\
6 & -580.99 & -596.7 & 16 & 308.11 & 301.0 & 7.1 \\
7 & -679.82 & -699.9 & 20 & 361.66 & 353.0 & 8.7 \\
    \hline
  \end{tabular*}
\end{table*}
\subsection{Polarizability due to electronic transitions }\label{theo:diss}
For static electric fields $(\omega\rightarrow 0)$, it is known that $\dpolrv\ \gg \dpole\ $\cite{Moss2002}. This may not necessarily be the case for infrared frequencies, for which $\dpolrv\ $ is expected to be smaller as spectrally nearby vibrational overtones are generally weak, whereas the detuning from strong rotational transitions and fundamental vibrations is large. Thus, there may be spectral regions where $\dpole\ $ becomes comparable in magnitude to $\dpolrv\ $. However, transitions from low-lying 1s$\sigma_g$ rovibrational states to 2p$\sigma_u$ states are located in the ultraviolet (UV) or even in the vacuum-ultraviolet (VUV) spectral range. Since the frequencies present in the $T=300$~K BBR spectrum are in the infrared (peak emission wavelength $\sim 10~\mu$m), it seems justified to regard the BBR electric field as static where it concerns $\dpole\ $. In Sec.~\ref{res:AccuracyApprox} it will be further justified that for this reason, $\dpoleO\ $ is a good approximation to $\dpole\ $.

Rather than deriving the static polarizability $\dpoleO\ $ from second-order perturbation theory, we extract its values from previously published and accurate static polarizabilities $\dpolrvO\ $, obtained by a full nonadiabatic calculation by Moss and Valenzano\cite{Moss2002}, as follows. From each of the total static polarizabilities $\dpolO\ $ tabulated by Moss and Valenzano, we subtract our value for $\dpolrvO\ $ calculated using the procedure described in Sec.~\ref{theo:rovib} to obtain $\dpoleO\ $.
\section{Results and discussion}\label{res}
\subsection{Rovibrational wavefunctions and dipole matrix elements}
Before discussing the results of our method to obtain $\dpol\ $, it will be worthwhile to investigate the accuracy of the wavefunctions $\chi_{vJ}(R)$ and energy levels $E_{vJ}$ obtained from Eq.~(\ref{eq:RadSchrod}), as well as the accuracy of the radial dipole matrix elements $\mu_{vv'JJ'}$ calculated using Eq.~(\ref{eq:radrobivdipmom}).
From comparisons with more accurate nonrelativistic level calculations for HD$^+$ \cite{Schiller2005} the inaccuracy of the energies $E_{vJ}$ calculated here is found to be a few parts in $10^5$ (or less than $0.5~$cm$^{-1}$), in correspondence with the accuracy specified by Esry and Sadeghpour\cite{Esry1999}. The accuracy of the energy levels also gives an indication of the accuracy of the wavefunctions $\chi_{vJ}(R)$.

In order to check the accuracy of the radial matrix elements $\mu_{vv'JJ'}$, a comparison can be made with values calculated by Colbourn and Bunker\cite{Colbourn1976}. Here it is important to note, however, that Colbourn and Bunker ignore effects of \textit{g/u} symmetry breaking by using a dipole moment function $D_{\mathrm{CB}}(R)\approx e R/6$ (with $e$ the electron
charge)\footnote[4]{This expression follows from evaluating the HD$^+$ 1s$\sigma_g$ dipole moment with respect to the center of mass at the equilibrium internuclear separation, for which the electron on average sits halfway the two nuclei.}. This functional form is valid at short internuclear range, where effects of \textit{g/u} symmetry breaking are small. However, for large internuclear separation in the 1s$\sigma_g$ state of HD$^+$, the electron sits primarily at the deuteron, which leads to a dipole moment function varying for large $R$ as $\approx (2/3) e R$. The function $D_1(R)$ provided by Esry and Sadeghpour includes effects of \textit{g/u} symmetry breaking, as illustrated in Fig.~\ref{fig:pots}(b). To compare with the results by Colbourn and Bunker, we first use our $\chi_{vJ}(R)$ with $D_\mathrm{CB}(R)$ to obtain matrix elements $\mu^\mathrm{CB}_{vv'JJ'}$. We find agreement at the level of a few times $10^{-5}$, consistent with the accuracy of both our wavefunctions $\chi_{vJ}(R)$ and those used by Colbourn and Bunker, which produce energy levels with similar accuracy. A second calculation using $D_1(R)$ instead of $D_\mathrm{CB}(R)$ leads to radial matrix elements differing from those by Colbourn and Bunker at the level of $2\times 10^{-3}$ for transitions $v'=1 - v=0$, and $4\times 10^{-3}$ for $v'=5 - v=4$. This difference we attribute to the inclusion of \textit{g/u} symmetry-breaking effects in $D_1(R)$, and may be considered an improvement over the values by Colbourn and Bunker. We put an conservative error margin of $25\%$ on this difference, thereby placing an upper bound of $1\times 10^{-3}$ on the accuracy of the matrix elements $\mu_{vv'JJ'}$.
\begin{figure*}[!ht]
  \centering
  \includegraphics[height=5.0cm]{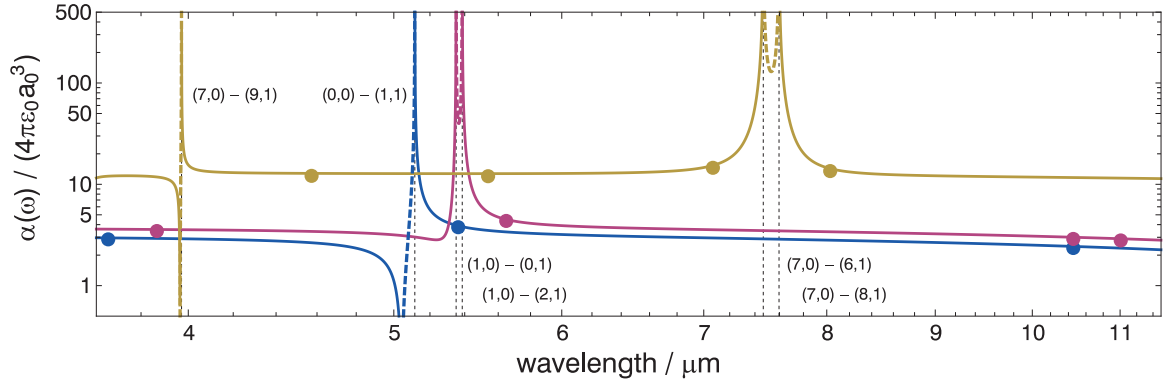}
  \caption{(Color online) Dynamic polarizabilities versus wavelength $(\lambda = 2\pi c/ \omega )$ for states with $J=0$ and, from bottom to top, $v=0,1,7$, respectively. The curves were produced using Eq.~(\ref{eq:dynpolresult}). For each curve,  dashed segments correspond to negative values, while solid segments correspond to positive values. Each 'dip' or 'peak' in a curve corresponds to a zero crossing of the polarizability. Vertical dashed lines indicate the position of rovibrational transitions $(v,J)-(v',J')$ coupling to $J'=1$ states. The dots show the more accurate values calculated for specific wavelengths by Karr \textit{et al.}\cite{Karr2005}, which agree with our result to within 3\%.}
  \label{fig:Karrpolplot}
\end{figure*}
\subsection{Static polarizability results}
\subsubsection{Accuracy of $\dpolrvO\ $.~~}
The results of Sec.~\ref{theo:rovib} enable us to calculate dynamic polarizabilities $\dpolrv\ $. To assess the accuracy of these calculations, we have checked the dependence of the static polarizability $\dpolrvO\ $ on the accuracy of both the energy levels and the radial matrix elements used. Computing $\dpolrvO\ $ once with the eigenvalues $E_{vJ}$ of Eq.~(\ref{eq:RadSchrod}), and once with accurate energy levels published by Moss\cite{Moss1993} (accuracy better than 0.001~cm$^{-1}$), we find that  $\dpolrvO\ $ varies by $\sim 1 \times 10^{-4}$. A similar check is done by using values $|\mu_{vv'JJ'}|^2$ computed using $D_\mathrm{CB}(R)$ and $D_1(R)$, respectively. The effect of the improved values on $\dpolrvO\ $  is a few times $10^{-3}$. Placing again a conservative bound of $25\%$ on the accuracy of this improvement, the accuracy of our value of $\dpolrvO\ $ is found to be $\le 1\times 10^{-3}$.
We also monitored the effect of the truncation of Eq.~(\ref{eq:Sobelmandef}) to $v'=9$. This has no noticeable effect at the $1\times 10^{-3}$ level for states $v \le 7$.
\subsubsection{Accuracy of $\dpoleO\ $.~~}
As described in Sec.~\ref{theo:diss}, the values $\dpolrvO\ $ may be combined with previously published values $\dpolO\ $ to extract $\dpoleO\ $. Thus-found values of $\dpoleO\ $ are presented in Tables~\ref{table:pols} and \ref{table:polsJ1M0}. We find that $\dpoleO\ $ contributes to $\dpolO\ $ at the $1\% $ level. Given the $\le 1\times 10^{-3}$ accuracy of our results for $\dpolrvO\ $, we are lead to believe that the values of $\dpoleO\ $ inferred here are accurate to within $10\%$.

 It is furthermore interesting to compare the values of $\dpoleO\ $ obtained here with static polarizabilities of the isotopomers H$_2^+$ and D$_2^+$, which were calculated with high accuracy for vibrational states with $J=0$ by Hilico \textit{et al.}\cite{Hilico2001}. In Table~\ref{table:comparison} it can be seen that for each vibrational state, the HD$^+$ value lies in between the values for H$_2^+$ and D$_2^+$. This is explained by the fact that the energy of a given vibrational state scales as $\sqrt{1/\mu}$, with $\mu$ the reduced nuclear mass of the isotopomer. Thus, for large reduced mass, vibrational levels are more deeply bound and therefore exhibit a smaller static polarizability. As the variation of binding energy is small compared to the typical energies of transitions to 2p$\sigma_u$ states, the mass scaling of the polarizability is approximately linear, and the value for HD$^+$ should be located halfway the values for H$_2^+$ and D$_2^+$ as in Table~\ref{table:comparison}.
\subsection{Dynamic polarizability results}\label{res:totdynpol}
\subsubsection{Accuracy of the approximation.~~}\label{res:AccuracyApprox}
As discussed in Sec.~\ref{theo:diss}, we will approximate the dynamic polarizability $\dpol\ = \dpolrv\ + \dpole\ $ by the expression
\begin{equation}\label{eq:dynpolresult}
\dpol\ \approx  \dpolrv\ + \dpoleO\ .
\end{equation}
For the infrared spectral range of interest here ($\lambda \ge 4~\mu$m) we believe that by approximating $\dpole\ $ by $\dpoleO\ $ we systematically underestimate the magnitude of the shift due to $\dpole\ $ alone by less than $10\%$ (details of this estimate are postponed to the Appendix). This is comparable to the uncertainty of the values $\dpoleO\ $ reported in Tables~\ref{table:pols} and \ref{table:polsJ1M0}. In order to verify the accuracy, we compare the result of Eq.~(\ref{eq:dynpolresult}) with the more accurate values calculated by Karr \textit{et al.} for a discrete set of wavelengths for states with $J=0$ (Fig.~\ref{fig:Karrpolplot}). The results of the two methods are found to agree within 1\% for $v=0$ and within 3\% for $v=7$. As the comparison is made for relatively short wavelengths, for which the polarizability stems almost entirely from $\dpole\ $, the level of agreement is consistent with the estimated error of $\le 10$\% in the value of $\dpoleO\ $.

The result for $\dpoleO\ $ obtained here is more useful than one would expect on the basis of its error margin for two reasons. First, for dynamic Stark shifts due to BBR (found by integrating the dynamic Stark shift over the BBR electric field spectral density; see Eq.~(\ref{eq:BBRshift}) and the Appendix), we estimate the error introduced by the quasi-static approximation to be even smaller, $\le 3\% $.  Second, for spectroscopy one is primarily concerned with differential level shifts, for which the systematic errors in $\dpol\ $ will partially cancel.
\begin{figure*}[!ht]
  \centering
  \includegraphics[height=7.8cm]{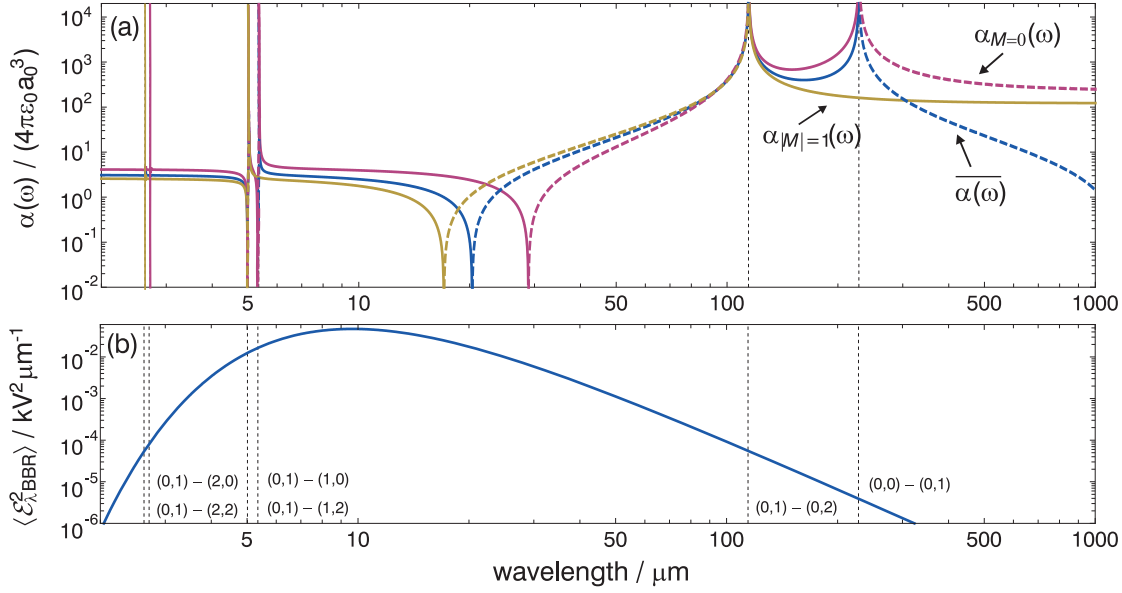}
  \caption{(Color online) (a) Dynamic polarizabilities versus wavelength $(\lambda = 2\pi c/ \omega )$ for various states with $v=0, J=1$, computed using Eq.~(\ref{eq:dynpolresult}). For each curve,  dashed segments correspond to negative values, while solid segments correspond to positive values. Each 'dip' or 'peak' in a curve corresponds to a zero crossing of the polarizability. Vertical dashed lines indicate the position of rovibrational transitions $(v,J)-(v',J')$ coupling to $J'=0,2$ states. The curves show marked tensorial differences between different $M$-states for polarized electric fields. It is also seen that for shorter wavelengths the contribution by rovibrational transitions becomes less significant, and that the electronic contribution becomes dominant instead. Furthermore, the magnitude of the average polarizability is seen to decrease towards longer wavelengths, which can be interpreted as an geometric averaging effect of the molecular rotation. (b) Mean-square electric field spectral density of the BBR at $T=300$~K. The BBR spectrum encompasses several rovibrational transitions, which implies that the Stark effect due to BBR is dynamic. Furthermore, the BBR spectrum covers both the rovibrationally-dominated (long-wavelength) polarizability range and the electronically-dominated (short-wavelength) range. This illustrates the need to include both rovibrational and electronic polarizabilities in a calculation of dynamic Stark shifts due to BBR.}
  \label{fig:dynpolplot}
\end{figure*}
%New paragraph starting below, so leave empty line intact!
\subsubsection{Dependence on $|M|$ and polarization state.~~} \label{res:EffectMPol}
It was mentioned in Sec.~\ref{theo:rovib} that Eq.~(\ref{eq:dynpolresult}) tacitly assumes linearly polarized electric fields. For obtaining the shift due to unpolarized, incoherent BBR, it is necessary to average over the three independent polarization states $q=-1,0,1$. It may be shown from Eq.~(\ref{eq:robivdipmom}) that this is equivalent to averaging Eq.~(\ref{eq:dynpolresult}) over all $M$ states:
\begin{equation}\label{eq:dynpolresultAv}
\dpolAv\  = \frac{1}{(2J+1)}\sum_M \dpol\ ,
\end{equation}
leading to a shift $\Delta E^\mathrm{BBR}_{vJ}(T)$ due to the BBR mean-square electric field density $\langle \mathcal{E}_\mathrm{BBR}^2(\omega,T)\rangle $ of
\begin{equation}
\Delta E^\mathrm{BBR}_{vJ}(T) = -\frac{1}{2} \int_{0}^{\infty} \dpolAv\ \, \langle \mathcal{E}_\mathrm{BBR}^2(\omega, T)\rangle \, \mathrm{d}\omega .
\label{eq:BBRshift}
\end{equation}
In our model, $\dpolAv\ $ involves a summation over terms which diverge for frequencies equal to their respective rovibrational transition frequencies (Eqs.~(\ref{eq:dynpol}) and (\ref{eq:Sobelmandef})). The integration over this sum is performed as follows. First, the convergence properties of the sum and BBR density function (Eq.~(\ref{eq:BBRdensity}) in the Appendix) allow to interchange the summation and integral signs, after which Eq.~(\ref{eq:BBRshift}) is evaluated as a series of Cauchy principal value integrals.

\begin{figure}[!ht]
  \centering
  \includegraphics[height=5cm]{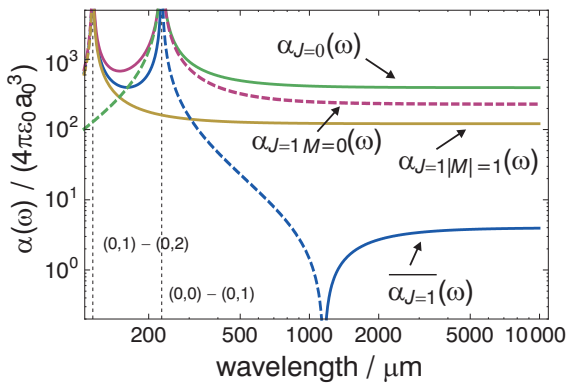}
  \caption{(Color online) Long-range wavelength behavior of the $v=0, J=1$ polarizabilities shown in Fig.~\ref{fig:dynpolplot}(a). Dashed segments of each curve correspond to negative-valued polarizabilities, solid segments to positive values. Vertical dashed lines indicate the position of rovibrational transitions $(v,J)-(v',J')$ coupling to $J=0,2$ states. In addition, the polarizability of the ($v=0, J=0$) state is shown, which is strictly scalar. Due to the absence of rotation, for this state the average polarizability due to rovibrational transitions does not average out as for $J=1$ states.}
  \label{fig:plotlongwl}
\end{figure}
\begin{table*}
\small
  \caption{\ Dynamic Stark shifts (in mHz) due to $T=300$~K BBR for various vibrational states with $J=0,1$ }
  \label{table:BBRshifts}
  \begin{tabular*}{\textwidth}{@{\extracolsep{\fill}}lllllll}
    \hline
     & \multicolumn{3}{c}{$J=0$} & \multicolumn{3}{c}{$J=1$}\\
    \cline{2-4} \cline{5-7}
    $v$ & Contribution $\dpolrvAv\ $ & Contribution $\dpoleAv\ $ & Total & Contribution $\dpolrvAv\ $ & Contribution $\dpoleAv\ $ & Total \\
    \hline
    0 & 35 & -27 & 8.3 & 32 & -27 & 4.6 \\
    1 & 38 & -33 & 5.5 & 34 & -32 & 1.9 \\
    2 & 41 & -39 & 1.6 & 37 & -39 & -1.9 \\
    3 & 43 & -47 & -3.5 & 40 & -47 & -7.1 \\
    4 & 46 & -57 & -11 & 43 & -57 & -14 \\
    5 & 49 & -70 & -21 & 46 & -70 & -24 \\
    6 & 52 & -81 & -29 & 49 & -86 & -37 \\
    7 & 55 & -111 & -56 & 52 & -107 & -55 \\
    \hline
  \end{tabular*}
\end{table*}
We stress that the average polarizability (Eq.~(\ref{eq:dynpolresultAv})) can be applied to unpolarized, incoherent electric fields only. To illustrate this, we plot (for $v=0$ and $J=1$) both the average polarizability $\dpolAv\ $ and the polarizabilities for linearly polarized electric fields $\dpol\ $ and $|M|=0,1$ in Fig.~\ref{fig:dynpolplot}(a). For long wavelengths, $\dpol\ $ is dominated by purely rovibrational transitions. This contrasts the situation for $\dpolAv\ $, in which the rovibrational contributions to the polarizability average out due to the molecular rotation (see also Fig.~\ref{fig:plotlongwl}). Several rotational and vibrational transitions occur which decay by spontaneous emission (spontaneous lifetime $\sim 10$~ms\cite{Amitay1994}). Hyperfine structure (which is ignored in our model) of these transitions covers a spectral range of about 1~GHz\cite{Bakalov2011}, which would not be visible on the scale of Fig.~\ref{fig:dynpolplot}(a). For wavelengths shorter than $20~\mu$m, electronic transitions start to dominate the dynamic polarizability, except for narrow spectral regions near vibrational transitions where the rovibrational polarizability diverges. Another remarkable feature is the absence of certain divergences in the $J=1, |M|=1$ polarizabilities which do appear in the $J=1, M=0$ polarizability. This is due to the selection rule $M'-M=0$ appertaining to electric fields linearly polarized along the $z$-axis (as assumed here). As a consequence, states with $J=1, M=0$ are coupled to states with $J'=0$, whereas states with $J=1, |M|=1$ are not, which explains the absence of $J'=0 - J=1$ divergences for $|M|=1$ polarizabilities. As expected, the average polarizability $\dpolAv\ $ contains all divergences.

Figure~\ref{fig:plotlongwl} shows the behavior of the dynamic polarizabilities of $J=0,1$ states at very long wavelengths (electric field frequency approaching dc). Here, it is clearly visible that the 'rotationless' $J=0$ state has large polarizability as there is no averaging effect by the rotation. In general, the dynamic polarizabilities display strong tensorial behavior, in particular in cases where the electric field is polarized. This is an important feature to bear in mind if Stark shifts due to the radio-frequency electric fields used in ion traps are to be considered, as these fields have a well-defined polarization.
\subsubsection{Results for BBR shift.~~}
As is obvious from Fig.~\ref{fig:dynpolplot}(b), the Stark effect due to BBR at $T=300$~K is dynamic. This situation differs radically from that for atomic ions, for which the Stark effect due to BBR radiation can be often treated quasi-statically. Thus, the treatment of systematic shifts in spectroscopy of HD$^+$ must be done with extra care, despite the fact that QLS of HD$^+$ molecular ions in the Lamb-Dicke regime may be done in a similar way as for atomic ions\cite{Schmidt2006long, Koelemeij2007a}.

Dynamic Stark shifts due to $T=300$~K BBR to several rovibrational levels are calculated by numerical integration of Eq.~(\ref{eq:BBRshift}) using the Cauchy principal value package of the \textit{Mathematica} computational program. Results are tabulated in Table~\ref{table:BBRshifts}, in which we also specify the individual rovibrational and electronic contributions. The rovibrational contributions turn out to produce positive level shifts. This can understood qualitatively from Figs.~\ref{fig:dynpolplot}(a) and (b). Indeed, the BBR spectrum samples primarily the rovibrationally-dominated spectral region ($\lambda > 20~\mu$m) where the polarizability attains negative values, leading to a positive level shift by virtue of Eq.~(\ref{eq:dynpol}). On the other hand, BBR wavelengths below $20~\mu$m primarily polarize the electronic structure of the molecule for which the polarizability is positive, and which explains the negative shift introduced by the electronic contribution (Table~\ref{table:BBRshifts}).  We also calculate differential BBR shifts to several transitions which may be amenable to Lamb-Dicke spectroscopy (Table~\ref{table:BBRdifshifts}). For optical transitions, the differential shifts are relatively small and contribute at the level of $10^{-16}$. Assuming that the temperature of the BBR field in an experimental apparatus\cite{Chou2010a} can be determined to within $\pm 10$~K, we find from Eq.~(\ref{eq:BBRshift}) that the BBR shift to optical transitions can be inferred from the polarizabilities derived here with relative accuracy better than 40\%, or well below $10^{-16}$ relative to the transition frequency. It should be noted that the shifts are much smaller than both the HD$^+$ hyperfine splittings\cite{Bakalov2006} and Zeeman shifts due to magnetic fields typically encountered in experiments\cite{Koelemeij2007a}. A more refined analysis of BBR shifts should therefore include the Zeeman effect as well as the hyperfine structure.
\section{Conclusion}\label{conc}
The dynamic polarizability of rovibrational states in the 1s$\sigma_g$ electronic state of HD$^+$ has been evaluated by combining  existing data on static polarizabilities with numerical calculations done using a simplified model of the HD$^+$ molecule. As a result of these numerical calculations, new values for radial dipole transition matrix elements were obtained which can be regarded as an improvement over existing values\cite{Colbourn1976}. The thus found dynamic polarizabilities point out that the Stark effect due to BBR -- an important systematic effect in optical spectroscopy of atomic ions and optical clocks -- is highly dynamic for the molecular ion HD$^+$, in contrast to BBR shifts to optical transitions in atomic ions\cite{Rosenband2006}. In this respect, the case of HD$^+$ is similar to that of neutral molecules\cite{Vanhaecke2007}. It is furthermore pointed out that the sign and magnitude of infrared dynamic polarizabilities depend strongly on the polarization state of the electric fields present. This insight is important for the evaluation of another well-known systematic shift in high-resolution spectroscopy of trapped ions, namely the Stark shift due to the trapping electric fields\cite{Berkeland1998long}. Notwithstanding these salient features of the HD$^+$ polarizability, it is shown that $T=300$~K BBR shifts become important for optical spectroscopy of HD$^+$ only at the $10^{-16}$ level. The smallness of the BBR level shifts furthermore suggests that future, more refined polarizability calculations should take magnetic-field interactions and hyperfine structure into account.
 %For footnotes in the main text of the article please number the footnotes to avoid duplicate symbols. e.g.  \footnote[num]{your text} the corresponding author \ast counts as footnote 1, ESI as footnote 2, e.g. if there is no ESI, please start at [num]=[2], if ESI is cited in the title please start at [num]=[3] etc. Please also cite the ESI within the main body of the text using \dag.
\begin{table*}
\small
  \caption{\ Differential dynamic Stark shifts (mHz) due to BBR at $T=300$~K for various rovibrational transitions }
  \label{table:BBRdifshifts}
  \begin{tabular*}{\textwidth}{@{\extracolsep{\fill}}llllll}
    \hline
    $(v',J') - (v,J)$ & Wavelength/$\mu$m & Contribution $\dpolrvAv\ $ & Contribution $\dpoleAv\ $ & Total & Relative /$10^{-16}$ \\
    \hline
    $(0,1) -(0,0)$  & 227.98 & -3.9 & 0.2  & -3.7 & -28.2\\
    $(1,0) -(0,1)$  & 5.3499 & 6.5  & -5.5 & 0.9  & 0.16\\
    $(4,1) - (0,0)$ & 1.4040 & 7.1  & -30  & -23  & -1.1\\
    $(4,0) - (0,1)$ & 1.4199 & 15   & -30  & -15  & -0.72\\
    \hline
  \end{tabular*}
\end{table*}
\appendix
\appendix \numberwithin{equation}{section}
\section{Appendix}
Here, we justify the approximations presented in Sec.~\ref{res:AccuracyApprox}. We start by noting that except the 2p$\sigma_u$ electronically excited state, all excited-state potential energy curves are located at large internuclear range, and that these excited states are connected to 1s$\sigma_g$ states by VUV transitions having very poor Franck-Condon overlap with 1s$\sigma_g$ states with low vibrational quantum number\cite{Esry1999}. Therefore, it is reasonable to assume that the larger part of $\dpole\ $ stems from bound-free transitions from 1s$\sigma_g$ to 2p$\sigma_u$, and that we can use the 2p$\sigma_u$ potential energy curve of Esry and Sadeghpour\cite{Esry1999} to estimate the effect of ignoring the dynamic part of the polarizability $\dpole\ $. To this end, we need to consider the dynamic Stark shift due to bound-free transitions. A bound state, subject to an oscillating electric field with photon energy $E=\hbar \omega$,  will undergo an energy shift $\hbar \Delta(E)$ due to off-resonant bound-free coupling, with the corresponding frequency shift being given by \cite{CohenTannoudji}
\begin{equation}\label{eq:PVintegral}
\Delta(E) = \frac{1}{ 2\pi } PV \int_{0}^{\infty} \frac{\Gamma(E')}{E-E'}\mathrm{d}E'.
\end{equation}
Here, \textit{PV} denotes the Cauchy principal value, which is evaluated numerically using the Cauchy principal value package of the \textit{Mathematica} computational program, and $\Gamma(E)/(2 \pi )$ stands for the bound-free transition rate (in s$^{-1}$) induced by an electric field with photon energy $E=\hbar \omega$. This transition rate can be obtained using Fermi's Golden Rule, an approach which was followed by Dunn\cite{Dunn1968} to calculate cross sections $\sigma_{vJ}(E)$ for photodissociation of H$_2^+$. These cross sections are proportional to bound-free radial matrix elements of the form
\begin{equation}
\sigma_{vJ}(E) \propto \frac{E}{\sqrt{E_f}}\,|\int_{0}^{\infty} \chi_{E_f J'}(R)D_{12}(R) \chi_{vJ}(R)\mathrm{d}R|^2,
\label{eq:PDcrosssection}
\end{equation}
where $\chi_{E_f J'}(R)$ represents a free (dissociating) state of nuclear motion in 2p$\sigma_u$ with asymptotic energy $E_f$. $E_f$ is related to $E$ and the dissociation energy $E^d_{vJ}$ of the bound state ($v,J$) by
\begin{equation}
E = E^d_{vJ}+E_f,
\end{equation}
where we have neglected the small (29~cm$^{-1}$) isotopic splitting between the 1s$\sigma_g$ and 2p$\sigma_u$ dissociation limits\cite{Esry1999}. It is important to note that the shape of $\Gamma(E)$ is governed by these wavefunctions via Eq.~(\ref{eq:PDcrosssection}), and that the 'dynamic' content of the shift $\Delta(E)$ is therefore determined by these wavefunctions.
We calculate $\chi_{E_f J'}(R)$ for the case of HD$^+$ by outward numerical integration of Eq.~(\ref{eq:RadSchrod}) for given energy $E_f$ while using the 2p$\sigma_u$ potential of Esry and Sadeghpour\cite{Esry1999}. We normalize the free-particle wavefunctions as done by Dunn\cite{Dunn1968}, after which they may be used to find photodissociation cross sections $\sigma_{vJ}(E)$ for various states with $v=0-7$ and $J=0,1$. These cross sections are averages over $M$ levels and therefore suited for a treatment of the shift due to BBR (Sec.~\ref{res:EffectMPol}). Multiplying $\sigma_{vJ}(E)$ with the flux of photons from the radiation electric field yields the transition (photodissociation) rate $\Gamma_{vJ}(E)$ of state $(v,J)$:
\begin{equation}
\Gamma_{vJ}(E)=2\pi \sigma_{vJ}(E) \frac{I}{\hbar \omega} = 2\pi \sigma_{vJ}(E) \frac{c \epsilon_0 \langle \mathcal{E}^2 \rangle }{E} .
\label{eq:Gamma}
\end{equation}
Here, we used the definition of the irradiance $I = c \epsilon_0 \langle \mathcal{E}^2 \rangle$. Inserting $\Gamma_{vJ}(E)$ into Eq.~(\ref{eq:PVintegral}) subsequently produces the level shift $\Delta_{vJ}(E)$.

To test the validity of the approximations made in Sec.~\ref{res:AccuracyApprox} we apply Eq.~(\ref{eq:PVintegral}) to two cases. In the first case, we adopt the approximation of Sec.~\ref{res:AccuracyApprox} by first deriving the mean-square value of the BBR electric field, $\langle \mathcal{E}_\mathrm{BBR}^2(T)\rangle$, inserting it into Eq.~(\ref{eq:Gamma}), and subsequently calculating the level shift in the limit that $E\rightarrow 0$ (\textit{i.e.} assuming a static field). In the second case, we obtain the level shift by proper integration of Eq.~(\ref{eq:PVintegral}) over the BBR energy spectral density.

For the first case, we find $\langle \mathcal{E}_\mathrm{BBR}^2(T)\rangle$ from the equation
\begin{equation}\nonumber
\frac{1}{2}\epsilon_0 \langle \mathcal{E}_\mathrm{BBR}^2(T) \rangle = \frac{1}{2}W(T),
\end{equation}
noting that only half of the integrated BBR energy density, $W(T)$, is stored in the electric field. $W(T)$ is found by integrating the BBR energy spectral density $w(\omega,T)\mathrm{d}\omega$:
\begin{eqnarray}
W(T)& =& \int_{0}^{\infty} w(\omega,T) \mathrm{d}\omega \nonumber \\
    &=& \frac{\hbar}{\pi^2 c^3}\int_{0}^{\infty}  \frac{\omega^3}{e^{\frac{\hbar \omega}{k_B T}}-1} \mathrm{d}\omega \nonumber \\
    &=& \frac{\pi^2 \left( k_B T \right) ^4}{15 \left( \hbar c \right) ^3}.
\label{eq:BBRdensity}
\end{eqnarray}
Inserting $\langle \mathcal{E}_\mathrm{BBR}^2(T)\rangle$ into Eq.~(\ref{eq:Gamma}), and inserting the resulting transition rate $\Gamma_{vJ}(E,T)$ into Eq.~(\ref{eq:PVintegral}), we obtain the quasi-static approximation to the frequency shift $\Delta^\mathrm{static}_{vJ,\mathrm{BBR}}(T)$ as
 \begin{equation}
\Delta^\mathrm{static}_{vJ,\mathrm{BBR}}(T) = \lim_{E \rightarrow 0 } \Delta_{vJ}(E,T). \nonumber
\end{equation}

For the second case, we rewrite the BBR mean-square electric field spectral density as
\begin{eqnarray}
\langle \mathcal{E}_\mathrm{BBR}^2(\omega,T)\rangle \, \mathrm{d}\omega &=& \frac{1}{ \epsilon_0}w(\omega,T)\mathrm{d}\omega \nonumber \\
&\equiv& \frac{1}{\hbar \epsilon_0}\tilde{w}(E,T)\mathrm{d}E.
\label{eq:EBBRdE}
\end{eqnarray}
After inserting Eq.~(\ref{eq:EBBRdE}) into Eq.~(\ref{eq:Gamma}) we obtain the spectrally integrated dynamic BBR shift $\Delta_{vJ,\mathrm{BBR}}^\mathrm{dyn}(T)$ upon evaluating the expression
\begin{equation}
\Delta_{vJ,\mathrm{BBR}}^\mathrm{dyn}(T) = \frac{c}{\hbar} PV \int_{0}^{\infty} \int_{0}^{\infty} \frac{\sigma_{vJ}(E')\, \tilde{w}(E',T)}{E'(E-E')}\mathrm{d}E' \mathrm{d}E.
\end{equation}
\balance
The errors introduced by the approximation in Sec.~\ref{res:AccuracyApprox} can now be simply evaluated from the ratio $\Delta^\mathrm{static}_{vJ,\mathrm{BBR}}(T) / \Delta_{vJ,\mathrm{BBR}}^\mathrm{dyn}(T)$ for various states $(v,J)$ and temperatures $T$. This is possible even so Eq.~(\ref{eq:PDcrosssection}) is incomplete; any numerical prefactor missing there will be common to both methods to compute $\Delta_{vJ,\mathrm{BBR}}(T)$, and cancel out in the ratio. For states with $v\le 7$, we find that the ratio $1-
\Delta_{vJ,\mathrm{BBR}}^\mathrm{stat}(T)/\Delta_{vJ,\mathrm{BBR}}^\mathrm{dyn}(T) \le 0.03$. Comparing shifts due
 \vspace{2mm}
 to monochromatic fields in a similar fashion, we observe that that the ratio $1- \Delta_{vJ}(0)/\Delta_{vJ}(E) \le 0.1$ for $\lambda = 4~\mu$m, and decreases to 0 in the static-field limit. This translates directly to the accuracy of $\dpol\ $ claimed in Sec.~\ref{res:AccuracyApprox}.

%\balance


\footnotesize{
\bibliography{110415Koelemeij_V2} %your .bib file
\bibliographystyle{rsc} %the RSC's .bst file
}

\end{document}